\documentclass[smallextended]{svjour3}

\usepackage{graphicx} 
\usepackage{psfrag} 
\usepackage{times}
\usepackage{mathrsfs} 
\usepackage{color} 
\usepackage{comment}
\usepackage{amsmath,amssymb}
\DeclareSymbolFontAlphabet{\mathrsfs}{rsfs}
\DeclareMathAlphabet{\mathcal}{OMS}{cmsy}{m}{n}

\newcommand{\scri}{\mathrsfs{I}} 
\newcommand{\be}{\begin{equation}} 
\newcommand{\ee}{\end{equation}}

\begin{document}


\title{Intermediate behavior of Kerr tails}
\author{An\i l Zengino\u{g}lu \and Gaurav Khanna \and Lior M.~Burko}
\institute{A.~Zengino\u{g}lu \at Theoretical Astrophysics, California Institute of
  Technology, Pasadena, California USA \and 
  G.~Khanna \at Department of Physics, University of Massachusetts, 
  Dartmouth, Massachusetts USA \and
  L.~M.~Burko \at Universit\'{e} d'Orl\'{e}ans, Observatoire des Sciences de l'Univers en Region Centre, LPC2E Campus CNRS, 45071 Orl\'{e}ans, France \\ and  
  Department of Physics, Chemistry, and Mathematics, Alabama A\&M University, Normal, Alabama 35762, USA \\ and
  Theiss Research, La Jolla, California 92037, USA}
\maketitle

\begin{abstract}
  The numerical investigation of wave propagation in the asymptotic
  domain of Kerr spacetime has only recently been possible thanks to the 
  construction of suitable hyperboloidal coordinates. The asymptotics revealed
  an apparent puzzle in the decay rates of scalar fields: the late-time rates
  seemed to depend on whether finite distance observers are in the
  strong field domain or far away from the rotating black hole, an apparent 
  phenomenon dubbed `splitting.' We discuss far-field `splitting' in the full field and near-horizon `splitting' in certain projected modes using 
  horizon-penetrating, hyperboloidal coordinates. For either case we propose 
  an explanation to the cause of the `splitting' behavior, and we determine uniquely decay rates that previous studies found to be ambiguous or    immeasurable. 
  The far-field `splitting' is explained by competition between projected modes. 
  The near-horizon `splitting' is due to excitation of lower multipole modes 
  that back excite the multipole mode for which `splitting' is observed. In both 
  cases `splitting' is an  intermediate effect, such that asymptotically in time 
  strong field rates are valid at all finite distances. 
  At any finite time, however, there are three domains with different decay 
  rates whose boundaries move outwards during evolution.  We then propose a formula for the decay rate of tails 
  that takes into account the inter--mode excitation effect that we study. 
  \end{abstract}

\section{Introduction}

The response of a black hole to generic perturbations has been an active
topic of interest since the study of Regge and Wheeler in 1957 \cite{Regge:1957td}.
It is now well-known that black hole perturbations satisfy wave equations. Generic solutions
to such wave equations consist of three stages: an initial 
transient during which the evolution depends on details of initial data; exponentially 
decaying oscillations called quasinormal mode ringing; and a late-time polynomial decay dubbed 
the Price tail. In this paper, we focus on the late-time behavior during which the field decays as $t^{n}$, 
where $t$ denotes time measured by stationary observers and $n$ denotes the decay rate 
(\cite{Price:1971fb}), and specifically distinguish between intermediate albeit late time, and asymptotically late--time  behaviors.

The historical analyses of the Kerr decay rates are based on asymptotic expansion methods,
and have also been studied extensively. Theoretical work has provided the general expected 
decay rates of perturbations in Kerr spacetime \cite{Barack:1999st,Barack:1999ma,Hod:1999rx,hod-2000-61,Poisson:2002jz}. These rates have also been computed 
numerically \cite{Krivan:1999wh,Burko:2002bt,Tiglio:2007jp,Burko:2007ju,Burko:2010zj}.
Today, our understanding of late-time Kerr decay rates of spherical harmonic 
$Y_{\ell m}$ modes of the scalar field includes as an essential element the distinction 
between initial modes and projected modes generated by coupling. Denoting the initial 
mode number by $\ell'$ and the projected mode number by $\ell$, and focusing on azimuthal ($m=0$) modes, the asymptotic 
finite-distance decay rates of projected modes was found to be given by  $n^R=-(\ell'+\ell + 1)$ for $\ell<\ell'$ 
and $n^R=-(\ell'+\ell + 3)$ for $\ell\geq\ell'$ where $R$ is a finite value for the Boyer--Lindquist 
radial coordinate. 
Decay rates along null infinity differ from those at finite distances \cite{Gundlach:1993tp}.
The rates along null infinity of Kerr spacetime predicted by Hod (\cite{Hod:1999rx}) 
$n^{\scri^+}=-(\ell+2)$ for $\ell \geq \ell'$ and $n^{\scri^+}=-\ell'$ for $\ell\leq \ell'-2$. 
In summary:
\be \label{eq:rates}\quad
n^R = \left\{ \begin{array}{ll}
-(\ell'+\ell + 1) & \mathrm{for}\quad \ell<\ell'\,, \\
-(\ell'+\ell + 3) & \mathrm{for}\quad \ell\geq \ell' \end{array}\right.
\quad
n^{\scri^+} = \left\{ \begin{array}{ll}
-\ell' & \mathrm{for}\quad \ell\leq \ell'-2 \,, \\
-(\ell+2) & \mathrm{for} \quad \ell\geq \ell' \end{array}\right.
\ee
We revisit this formula in Section \ref{sec:inter} with particular focus on the cases $\ell'=\ell$, and 
our results suggest to us that Eq.~(\ref{eq:rates}) needs to be revised to include high--order couplings that 
are not captured in Eq.~(\ref{eq:rates}). 
The spherical harmonic modes ($\ell$,$m$) 
in this work are the same as those in earlier works that use Boyer--Lindquist 
coordinates \cite{Burko:2007ju,Burko:2010zj,Zenginoglu:2009hd} and other coordinates 
in the same equivalence-class \cite{Racz:2011qu,Jasiulek:2011ce}. Thus, the same 
late-time decay rates are observed for each mode.  

Until recently, the numerical study of Kerr tails in the literature has exclusively
focused on finite distance rates even though the null infinity rates are arguably more interesting 
for observers at astronomical distances (\cite{Purrer:2004nq,Zenginoglu:2008wc}).
Null infinity decay rates in Schwarzschild spacetime have been studied with the characteristic 
method as early as 1994 (\cite{Gundlach:1993tp}). The characteristic method, however, 
is difficult to extend to Kerr spacetime (\cite{Pretorius:1998sf,FletcherLun,Venter:2005cs}). 
It is due to this technical obstacle that Kerr tails in the asymptotic domain have not been 
studied numerically.

This problem has recently been resolved with the general construction of hyperboloidal
coordinates for asymptotically flat black hole spacetimes (\cite{Zenginoglu:2007jw}). 
These coordinates provide numerical access to the asymptotic domain and resolve 
the outer boundary problem in an efficient way. Recent hyperboloidal evolutions
(\cite{Zenginoglu:2009hd,Racz:2011qu,Jasiulek:2011ce,Harms:2013ib}) confirmed the theoretically 
predicted decay rates of projected modes at null infinity.

The hyperboloidal evolutions revealed an apparently puzzling behavior for the rates 
in the asymptotic spatial domain. The rates at finite distances far away from the 
black hole and at finite times appeared to deviate from the rates both near the black hole (including the 
event horizon), and along null infinity. This observation requires some 
explanation. In Schwarzschild spacetime, the local (in time) decay rates at finite 
distances depend on the distance to the black hole. For example, the dominant $\ell'=\ell=0$ 
mode decays along null infinity with a power of $-2$, and at finite distances with $-3$, 
asymptotically in time. Therefore, at any finite time, one expects that the finite distance 
rates vary monotonously between $-2$ and $-3$. This expectation is in accordance with 
theoretical predictions (see Eq.~(\ref{eq:pointwise}) and \cite{StraussTsutaya,Szpak:2008jv}) 
and is confirmed by numerical studies (\cite{Purrer:2004nq,Zenginoglu:2008wc}).

A similar distance-dependence for the full field is observed in Kerr spacetime for 
$\ell'<4$, but for initial data with $\ell'=4$ an anomalous behavior appears. 
Instead of a monotonous transition between the null infinity rate of $-4$ and the
finite distance rate of $-5$, there are apparently far away observers for whom the decay rate 
seems to approach a value smaller than $-5$ (\cite{Zenginoglu:2009hd}). 
A more detailed study of tail decay rates by R\'{a}cz and T\'oth in \cite{Racz:2011qu} provided independent evidence for 
such anomalous behavior. They suggested that certain projected modes have different 
local decay rates close to the horizon and far away from it, which they referred to 
as `splitting.' They also found that the `splitting' depends upon the value of 
the azimuthal mode number $m$. Numerical computations in 3D by Jasiulek showed that 
`splitting' is robust (\cite{Jasiulek:2011ce}). These studies pointed out that longer 
time evolutions are needed, suggesting that `splitting' may not be an asymptotic but rather an intermediate phenomenon. 
R\'{a}cz and T\'oth in \cite{Racz:2011qu} referred to certain decay rates as ``ambiguous," indicating that they were unable, 
because of the `splitting' phenomenon they observed, to determine uniquely the value of the decay rate $n$. 
More recently, Harms, Bernuzzi, and Bruegmann \cite{Harms_2013} revisited this question for scalar, electromagnetic, and gravitational 
perturbations and reported on similar 
`splitting' phenomena. They determined some of the rates that  \cite{Racz:2011qu} listed as ``ambiguous" but did not provide rates for all of them. 

In this paper, we perform careful evolutions that are nearly an order-of-magnitude longer 
than previous works (on the order of $10^4M$ where $M$ denotes the mass of the Kerr 
black hole), to study the two types of `splitting' discussed in the literature: 
`splitting' in the full field (\cite{Zenginoglu:2009hd}) and `splitting' in certain projected modes 
(\cite{Racz:2011qu,Jasiulek:2011ce,Harms_2013}). Our long evolutions, enabled by the advances in numerical technology, 
allow us to explain the origin of the two kinds of `splitting' phenomena, conclude that `splitting' does {\em not} exist in the asymptotically late--time decay rate but only as an intermediate feature of the fields evolution, and also determine uniquely all the decay rates that have been found previously to be ambiguous or immeasurable. We emphasize that our evolutions were just as long as needed to determine the new results: shorter evolutions might indeed lead one to conclude that some decay rates are ambiguous

We show that the two `splitting' phenomena arise from different mechanisms 
and argue that, asymptotically in time, the strong field 
decay rates are valid at all finite distances, such that {\em there is no `splitting' in the 
asymptotic regime}.  The full--field `splitting' has been explained before, yet we discuss it here again for the completeness of the presentation. 
`Splitting' in the full field is nearly--trivially explained as a competition between 
the constituent multipole modes, such that multipole modes which dominate at early times become eventually sub-dominant. 
`Splitting' in projected modes involves a more intricate mechanism: it appears as an 
intermediate behavior of decay rates from excitation of lower 
multipole modes that  back excite the higher mode in question. Both types of `splitting' disappear asymptotically. `Splitting' in the projected 
modes as an intermediate phenomenon is still interesting and important though, as 
at any given late time there are three domains observed with different local decay rates for 
certain projected modes. The boundaries between these domains 
move to larger distances during evolution. 

Our study requires integration into very late times. The technological advances that allowed for such long evolutions 
may be relevant also for other interesting problems. For example, 
the late-time decay of scalar fields is related to the violation of the Huygens' principle in black
hole spacetimes and therefore carries theoretical interest. Its presence indicates
that waves not only propagate on the light cone of the initial perturbation, but also within
the light cone. Such propagation is partly responsible for the late-time polynomial decay, 
also called the tail.
The tail piece of the solution is very small, and unlikely to be directly detectable. One may therefore
argue that it is not interesting from an astrophysical point of view. However, there are astrophysical 
problems where the accurate computation of the tail decay is important.

One specific astrophysical problem where one needs to numerically compute the 
late-time tail solution is related to the inspiral of a small black hole into a supermassive one. 
The self-force that acts on such a small black hole can be described by a 
time integral into the infinite past of the retarded Green function that describes perturbations of the 
supermassive black hole \cite{Mino:1996nk,Quinn:1996am}. 
The computation of such black hole Green functions at a given base point can be performed by
numerically integrating wave equations with vanishing initial data and a narrow Gaussian source
centered at the given base point approximating a Dirac distribution \cite{Zenginoglu:2012xe}. 
These computations must be performed into late times where the tail effect dominates because the 
self-force calculation includes a half-infinite time integral of the Green function  
\cite{Casals:2013mpa}. Therefore accurate computations of late-time wave solutions are important 
for a theoretical understanding of wave propagation and for
the development of numerical tools to tackle astrophysical problems.

The organization of this paper is as follows. In Section \ref{sec:methods} we describe the numerical methods that 
we use, specifically the hyperboloidal compactification of the horizon and of infinity in Section \ref{sec:layers} and the numerical 
implementation in Section \ref{sec:num}. In Section \ref{sec:inter} we describe our results for the far--field (\ref{far_field}) and the (near--field \ref{sec:nh}) cases. We discuss our results in Section \ref{sec:disc}.

\section{Methods}\label{sec:methods}

\subsection{Hyperboloidal compactification of the horizon and of infinity}
\label{sec:layers}

A technical new development in this paper is the application 
of the hyperboloidal layer method from \cite{Zenginoglu:2010cq} 
to the construction of horizon-penetrating, hyperboloidal coordinates in Kerr 
spacetime. The causal behavior of the slicing is similar to
the slicings of \cite{Racz:2011qu,Jasiulek:2011ce} with the difference that 
standard Boyer--Lindquist coordinates can be used in a compact domain. 
Our coordinates extend those in \cite{Zenginoglu:2011zz} to 
become horizon-penetrating. Similar coordinates have
been constructed previously for Schwarzschild spacetime in \cite{Bernuzzi:2012ku}. 

Denoting the tortoise coordinate based on the Boyer--Lindquist representation 
of the Kerr metric as $r_*$, we map the infinite domain $r_*\in(-\infty,\infty)$ 
to a finite domain $\rho\in[-S,S]$ using the spatial compactification
\be\label{eq:spat} r_* =\frac{\rho}{\Omega}, \quad \rm{with} \quad \Omega = 
1 - \left(\frac{|\rho| - R}{S-R}\right)^4 \Theta(|\rho| - R)\,, \ee 
where $\Theta$ denotes the step function and $R$ the location of the interface. 
The function $\Omega$ vanishes at $\rho=\pm S$ with a nonzero gradient. It is 
unity for $-R<\rho<R$ and sufficiently smooth at the interfaces $\rho=\pm R$. 
To avoid loss of resolution for outgoing waves near infinity, we combine the 
spatial compactification with a suitable time transformation introducing a new 
time coordinate $\tau = t \pm h(r_*)$. The function $h(r_*)$ is called the height 
function. 

The Kerr metric in Boyer--Lindquist coordinates is asymptotically
Schwarzschild in standard Schwarzschild coordinates. Therefore, the
leading order form of in- and outgoing null rays can be written as 
$t\pm r_*$. The height function is chosen such that the time function 
$\tau$ satisfies the condition $t\pm r_*=\tau\pm\rho$.
Combined with the spatial compactification (\ref{eq:spat}) we get 
\be\label{eq:heightBL} h(\rho) =  \frac{\rho}{\Omega} - \rho. \ee 
The choices (\ref{eq:spat}) and (\ref{eq:heightBL}) fully determine the 
coordinates. Note that only the asymptotic form of the in- and outgoing null 
surfaces goes into the hyperboloidal transformation.

\subsection{Numerical implementation}\label{sec:num}
We study local decay rates by solving the scalar wave equation, $\Box \phi = 0$,
for a rescaled variable $r \phi$, using hyperboloidal layers. Our numerical 
simulations are performed with a modified version of the time-domain evolution 
code presented in \cite{Burko:2010zj}. The code is an explicit, hyperbolic-PDE solver 
in (2+1)D that uses the Lax--Wendroff finite-difference evolution scheme. 
The numerical results have low truncation error due to high-order differencing 
in the angular direction and high-precision (quadruple or octal) floating-point 
operations. The code is parallelized via message passing interface using the standard 
domain-decomposition approach on the radial coordinate grid. The main modification 
of the code for this paper is the implementation of hyperboloidal layers as 
discussed in the previous section. 

Typical grid resolutions used in this work are $M/64$ in the radial and 
$\pi/48$ in the angular direction. We set $M=1$ and $a/M=0.995$. The domain 
size in the compactified $\rho$-coordinate is $[-100,100]\,M$ with the layer interfaces 
at $R=\pm 28M$. The time-step is set to $M/128$ as dictated by the Courant condition 
for stability of the numerical evolution scheme. The compact-supported, non-stationary, 
initial data for the scalar field is chosen to be a truncated Gaussian wave packet of 
width $4M$ and variable center location. This corresponds to type 1 initial data as 
classified in \cite{Racz:2011qu}. 

\section{Results}\label{sec:inter}
We investigate the decay rates for all $\ell'\leq 5$, specializing to vanishing azimuthal 
number, $m=0$. We study intermediate decay rates for both the projected modes and the full 
field measured by various observers using the local decay rate $n^{\rm{obs}}(t)$ defined 
as (\cite{Burko:1997}) 
\be \label{eq:rate} n^{\rm{obs}} := \frac{d\ln
  |\phi(t,r^{\rm{obs}})|}{d\ln t}. \ee 
We then define the asymptotic decay rate as $n_{\infty}^{\rm obs}:=\lim_{t\to\infty}n^{\rm obs}$.     
Note that the local and asymptotic decay rates are invariant under the transformations in the hyperboloidal 
layers because the timelike Killing field is left invariant by the hyperboloidal time transformation.

\subsection{Far-field}\label{far_field}
\subsubsection{Local rates of projected modes}
\begin{figure}[ht]
\includegraphics[width=0.5\columnwidth]{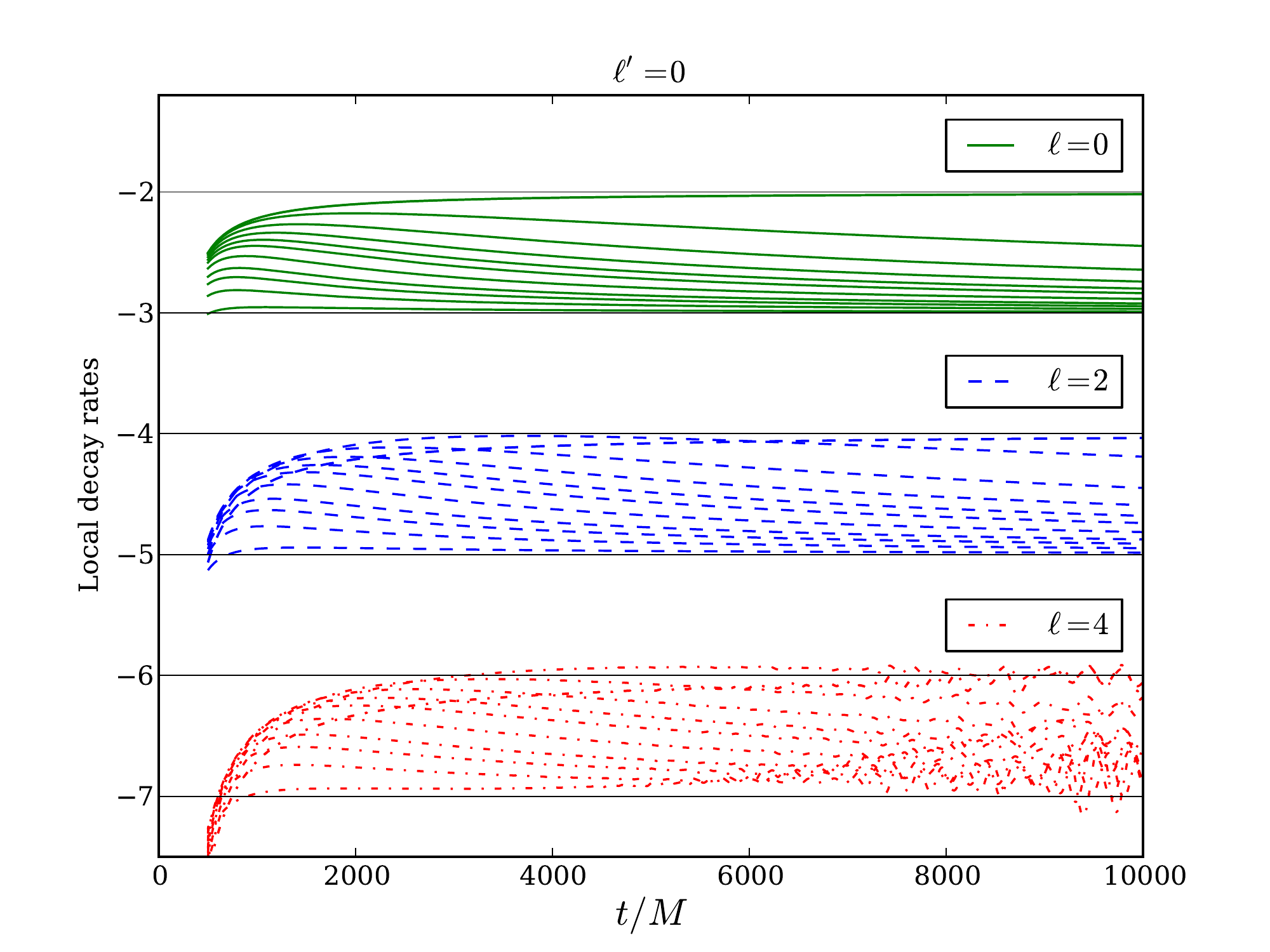}
\includegraphics[width=0.5\columnwidth]{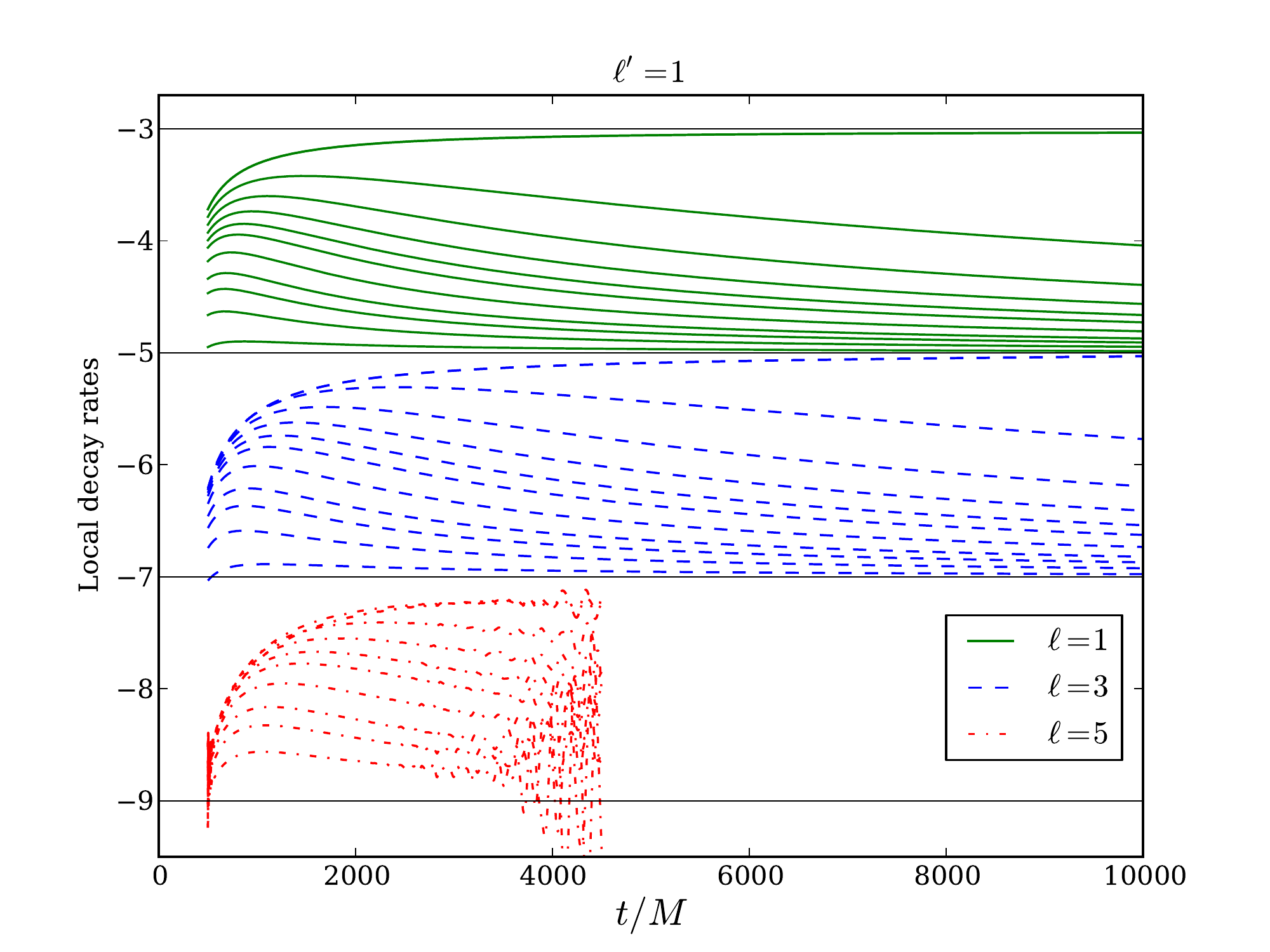}
\includegraphics[width=0.5\columnwidth]{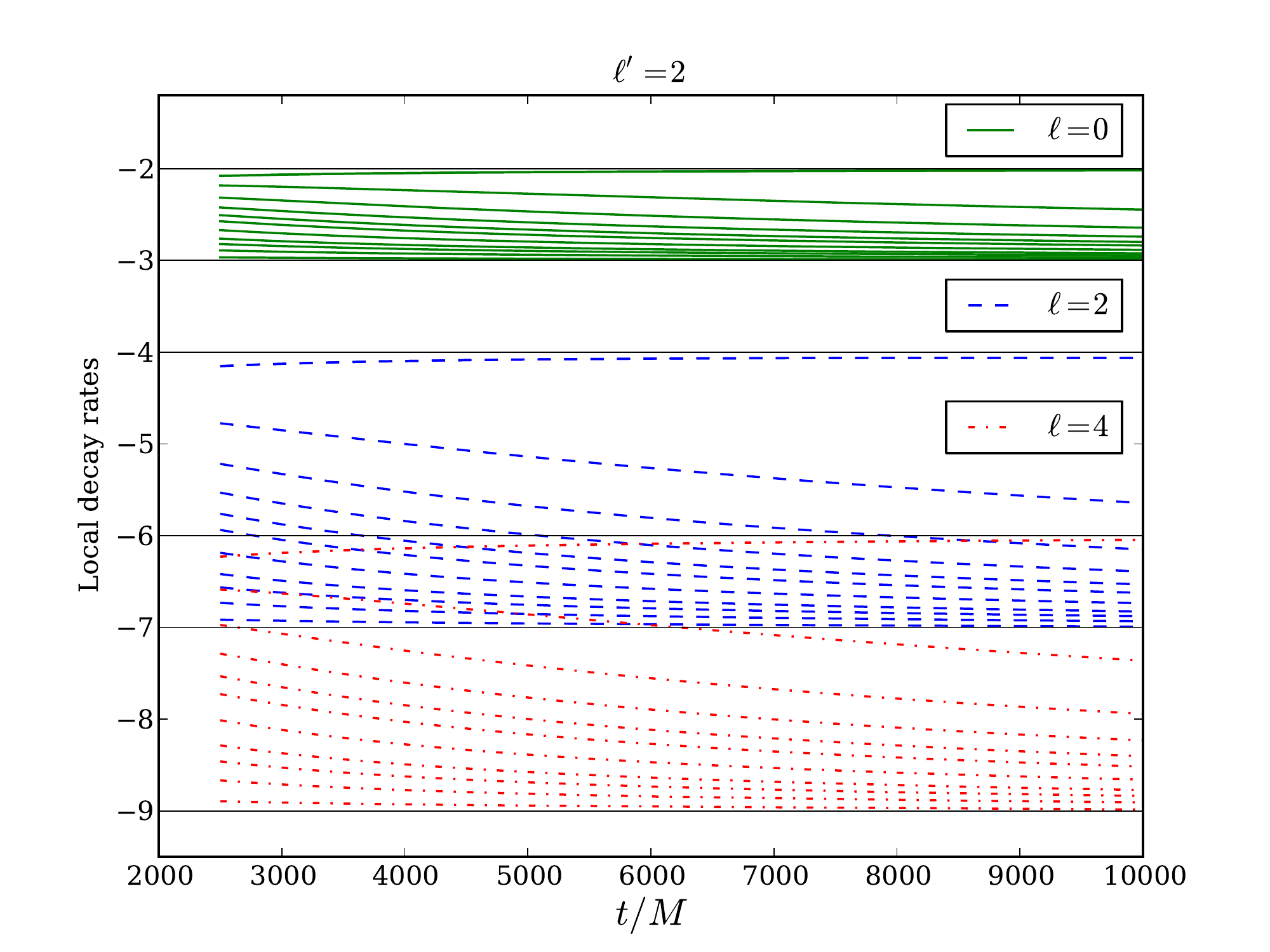}
\includegraphics[width=0.5\columnwidth]{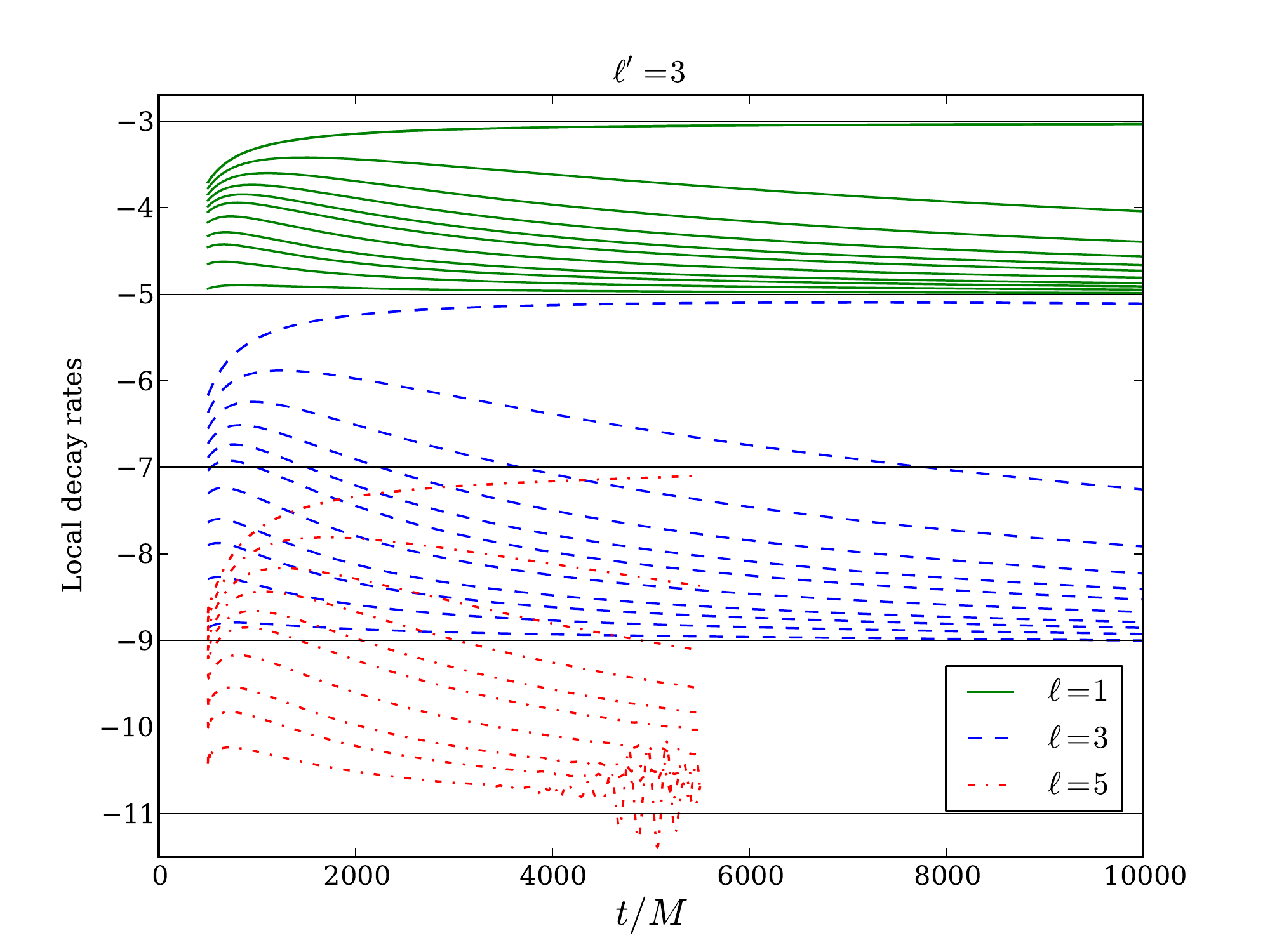}
\includegraphics[width=0.5\columnwidth]{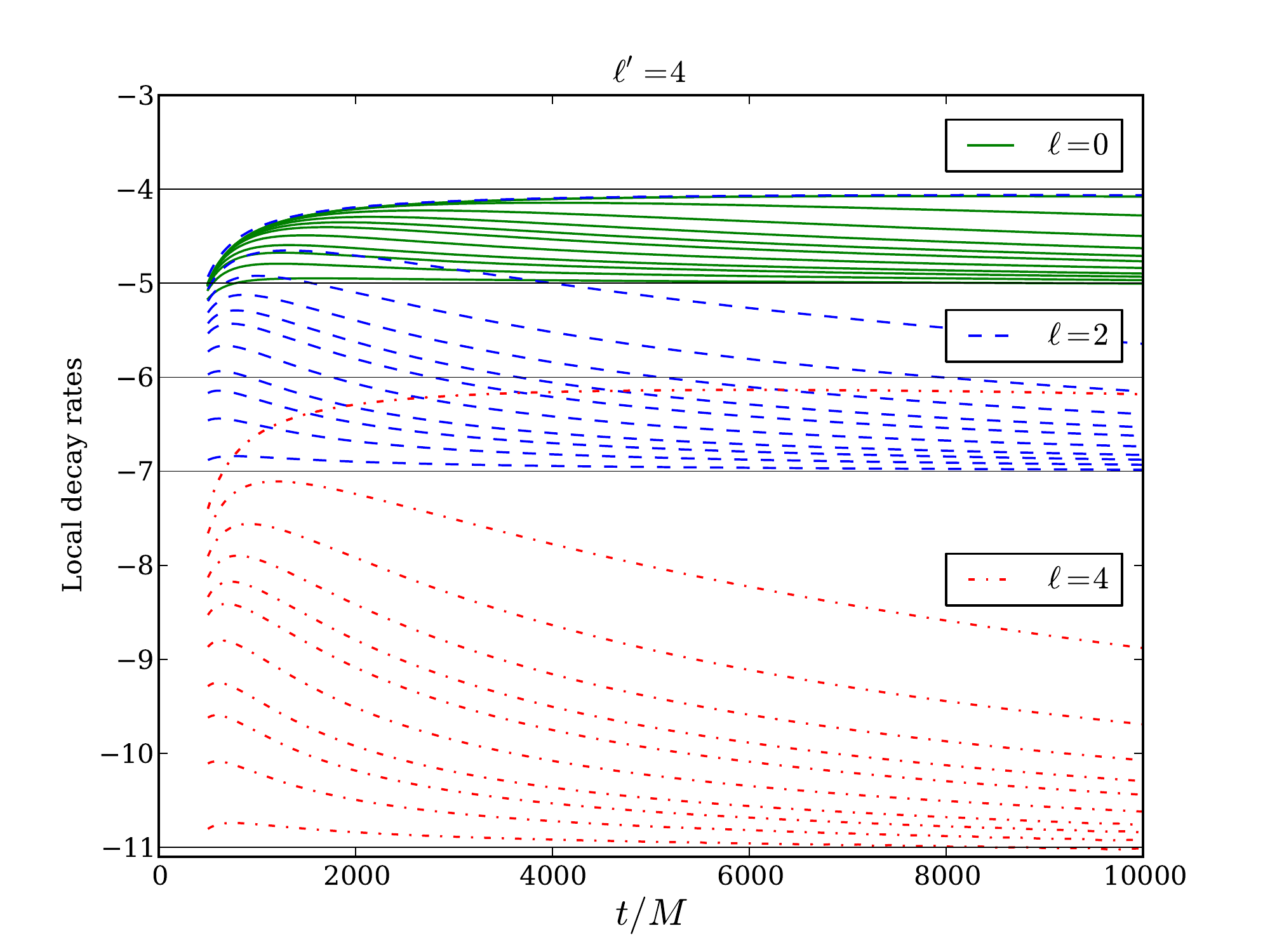}
\includegraphics[width=0.5\columnwidth]{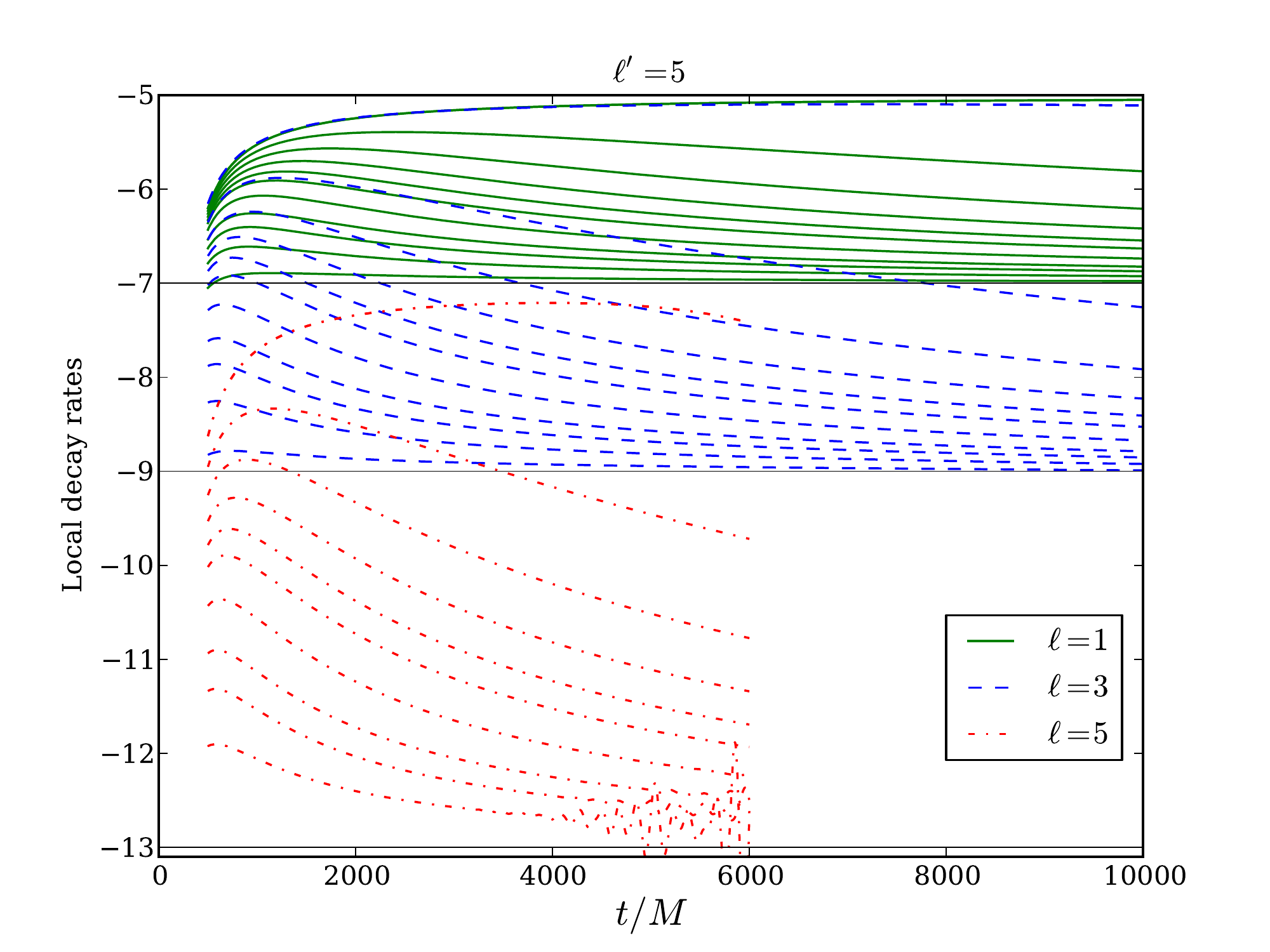}
\caption{Local decay rates for the evolution of even initial modes $\ell'=0,2,4$ 
on the left panels and odd initial modes $\ell'=1,3,5$ on the right panels, 
for a family of far-field observers ranging from null infinity to about $100M$ 
successively closer to the Kerr hole from top to bottom for each projection. 
The initial data is a Gaussian with compact support centered at $r_*=25M$. 
The late-time values are consistent with the predictions in the published literature. 
No `splitting' is observed in the far field ($r\geq 100M$) for $t=10^4M$.
The lowest mode rates in the odd modes (red dash-dotted on the right panel) are truncated due to numerical oscillations.
\label{ProjRates}}
\end{figure}

The intermediate decay rates, as defined in \eqref{eq:rate}. for projected modes with initial data mode $\ell' \leq 5$ 
measured by far-field observers ranging from null infinity to about $100M$ are shown in 
Fig.~\ref{ProjRates} with the center of the Gaussian initial perturbation located at $r_*=25M$. The decay rates are monotonous; no `splitting' occurs.

It is instructive to contrast Fig.~\ref{ProjRates} to observations in Schwarzschild 
spacetime. The observer dependence and the intermediate behavior of local decay
rates are simple there. An initial pure spherical harmonic stays a pure spherical 
harmonic during evolution ($\ell=\ell'$) and the decay rates are given as $n_{\infty}^{i^+}=-(2\ell+3)$ 
at timelike infinity and as $n_{\infty}^\scri=-(\ell+2)$ at null infinity \cite{Price:1971fb,Gundlach:1993tp}. 
During the intermediate decay there is a monotonous transition depending on the location 
of the observer (\cite{Purrer:2004nq,Zenginoglu:2008wc}). 

In Kerr spacetime, there is no geometric notion of a pure multipole 
because of the lack of spherical symmetry of the background spacetime: the scalar spherical harmonics are 
not eigenfunctions of the Laplace operator, and this leads to coupling between multipoles. 
Nevertheless, we observe a similar intermediate behavior in the projected modes for far 
field observers ($r>100M$) in Fig.~\ref{ProjRates}. 

The observer-dependence in these plots can be understood by considering the description 
of pointwise decay estimates suggested in Schwarzschild spacetime for small 
solutions (\cite{StraussTsutaya,Szpak:2008jv}). For the dominant
$\ell'=\ell=0$ mode the estimate in standard Schwarzschild coordinates reads
\be\label{eq:pointwise} |\phi| \leq
\frac{C_1}{(C_2+t+r_*)(C_3+t-r_*)^2},\ee where $C_i$ are
constants. This estimate captures the asymptotic behavior of the
solution near null infinity as well as at finite distances from the
black hole. It also provides an explicit description of the point- 
and slicing-dependence of the decay rates \cite{Zenginoglu:2008wc}. 
The powers in the above formula are different for different values of $\ell$, 
but the qualitative structure of observer-dependence is the same.

The far-field local-in-time decay rates suggested by Fig.~\ref{ProjRates} are 
listed in Tab.~\ref{tab:rates}. Notice that in Tab.~\ref{tab:rates} the entries 
that were indeterminate in Tab.~1 of \cite{Racz:2011qu} are printed with an asterisk. 
We emphasize that these are the decay rates that would be observed by late-time 
observers who are {\em not} asymptotic. Initial data (compact support data 
centered at $r_*=25M$) and observation points (far away observers with $r_*\geq 100M$) 
were chosen here so that even very long evolutions do not reveal the inter-mode 
interaction effect in the near field to be discussed below in Sec.~\ref{sec:nh}.

\begin{table}[h]
  \begin{center}
      \begin{tabular}{|c||c|c|c|}
	\hline
        $\ell'$   &   $\ell=0$  &    $\ell=2$   &   $\ell=4$ \\
        \hline\hline 
        0 & -2 \quad -3   &   -4\quad -5   &  -6\quad -7   \\
        2 & -2 \quad -3   &   -4\quad -7$^*$   &  -6\quad -9$^*$   \\
        4 & -4 \quad -5   &   -4\quad -7   &  -6\,\,\    -11$^*$ \\
	\hline
   \end{tabular}\qquad
      \begin{tabular}{|c||c|c|c|}
	\hline
        $\ell'$   &   $\ell=1$  &    $\ell=3$   &   $\ell=5$ \\
        \hline\hline 
        1 & -3 \quad -5   &   -5\quad -7   &  -7\, \quad  -9   \\
        3 & -3 \quad -5   &   -5\quad -9$^*$   &  -7\quad -11$^*$   \\
        5 & -5 \quad -7   &   -5\quad -9   &  -7\quad -13$^*$ \\
	\hline
   \end{tabular}
\end{center}
  \caption{Far-field decay rates for all mode projections suggested by Fig.~\ref{ProjRates} for even
   modes (left table) and odd modes (right table). We list both  $n_{\infty}^{\scri}$ (left entry of each
   column) and $n_{\infty}^{i^+}$ (right entry) for each choice of $\ell',\ell$. 
   The values agree with theoretical predictions listed in (\ref{eq:rates}) 
   (see \cite{Hod:1999rx,Burko:2010zj}). 
   However, some of the finite distance rates, marked with an asterisk, may change asymptotically 
   in time due to propagation of near-horizon inter-mode coupling behavior not seen 
   in Fig.~\ref{ProjRates} because the initial data and the observers are located far away 
   from the black hole (see Sec.~\ref{sec:nh}). 
  \label{tab:rates}}
\end{table}

\subsubsection{Local rates of the full field}
\begin{figure}[ht]
\center
\includegraphics[width=0.52\columnwidth]{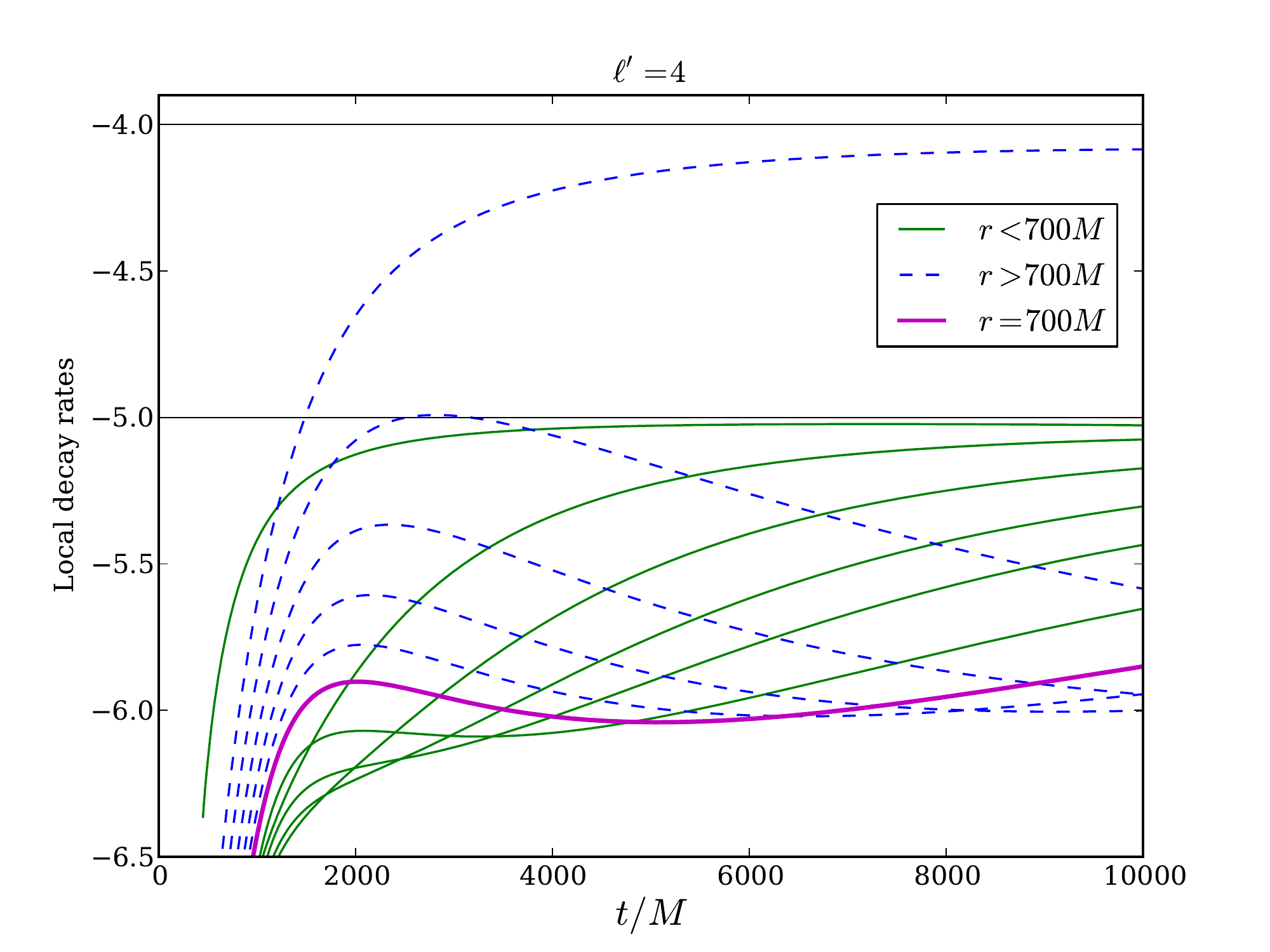}\hfill
\includegraphics[width=0.46\columnwidth]{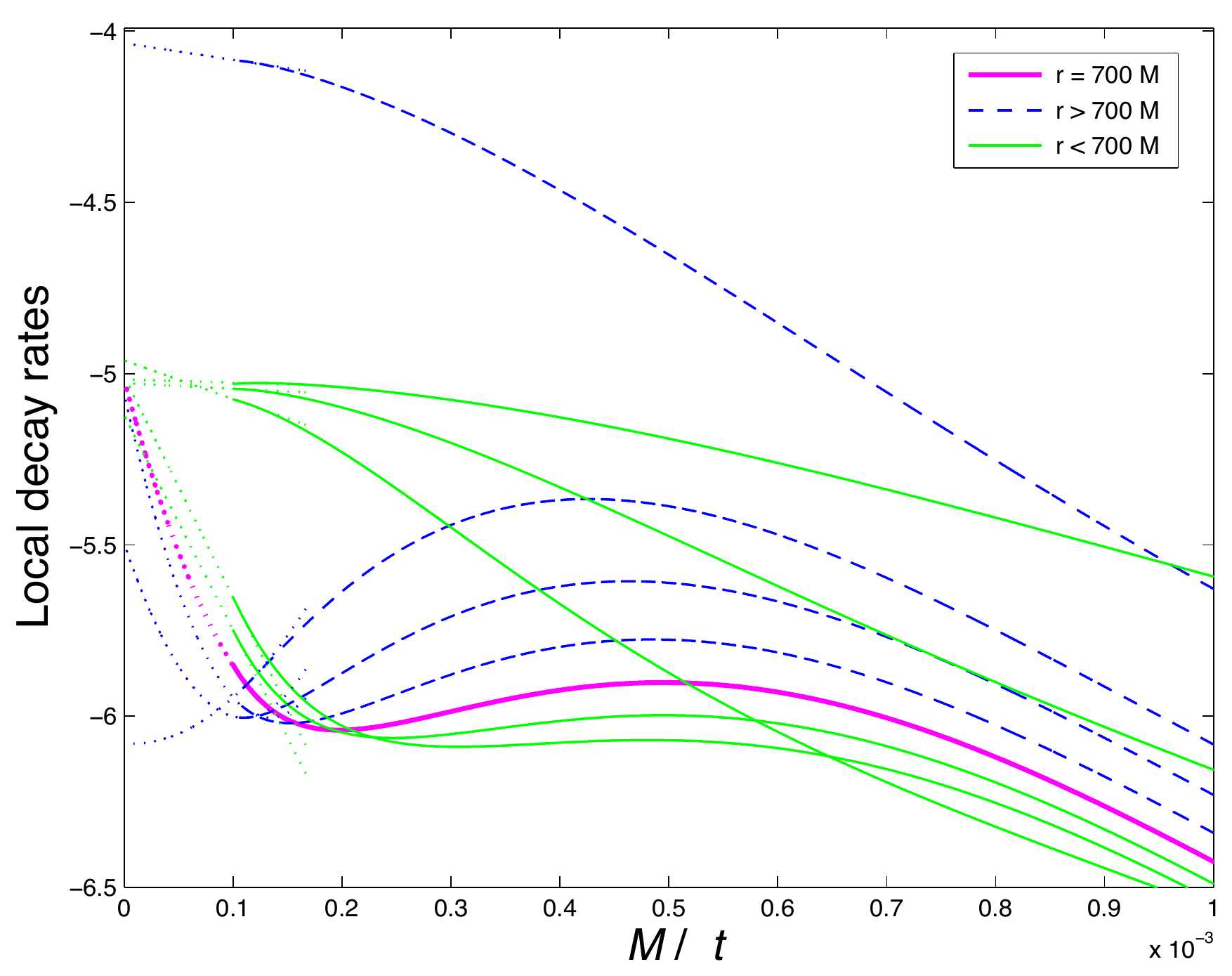}\hfill
\caption{The local decay rates for $\ell'=4$ initial data as functions of time ($t/M$, left panel) 
and inverse time ($M/t$, right panel). The bold solid magenta curve corresponds to an observer 
at $r=700M$, dashed blue curves for observers at $r>700M$, and thin solid grin curves to observers 
at $r<700M$. For the right panel, for high $M/t$ the distance of the observer increases from bottom 
to top. Note that the uppermost two thin solid green curves would appear below the thin solid green 
curves at the lower right corner were the figure extended to higher values of $M/t$. The dotted 
curves on the right panel are extrapolations of the local decay rates using linear extrapolation 
for $r<700M$ and quadratic extrapolations for $r\ge 700M$. 
\label{lpi_inv_time}}
\end{figure}

The `splitting' in the decay rates of the full field has first been observed in
\cite{Zenginoglu:2009hd}. We present in Fig.~\ref{lpi_inv_time} 
the local decay rate for the full field for initial data of $\ell'=4$ for a family of observers in linear (left panel) and inverse (right panel) time. 

The curves separate into two types: those corresponding to 
near observers for whom the local decay rate at the end point of the numerical simulation 
increases as time increases (or as inverse time decreases), and those corresponding to far 
observers for whom it decreases in the same limits. Notice that the designations `near' 
and `far' here are different than in the preceding section. The designation of near or far 
observers depends on how long the simulation runs. For our choice of $t_{\rm final}=10^4M$, 
the two types of curves separate at about $r\sim 700M$. For near observers ($r\lesssim 700M$) 
extrapolations of the local decay rate as a function of inverse time to $M/t\to 0$ produce 
results very close to the expected asymptotic value of $-5$, but far away observers seem to 
measure a decay rate of $-6$. Quantitatively, the observers and the extrapolated decay rates 
they measure are as follows: $(r_*/M,n_{\infty}^{i^+})=$ $\{ (112,-4.962), $ $(523, -5.008), $ $(608, -5.121), 
$ $(728,-5.017), $ $(907, -5.058), $ $(1207,-5.495), $ \\ $(1806, -6.08)\}$. 
The latter two values vary significantly from $-5$, which is what is referred to as the 
full-field `splitting.' Notice, however, that Fig.~\ref{lpi_inv_time} also shows how the `splitting' behavior is an intermediate one: 
the $(r_*/M,n_{\infty}^{i^+})=(1207,-5.495)$ case shows that although the extrapolated value deviates significantly form the 
expected value of $-5$, the curve already starts to curve up as $M/t$ decreases. The integration, however, is not long enough to allow for 
sufficient curvature to enter the curve, that would push the asymptotic value all the way up to $-5$. The situation would be the same if we only integrated any of the solid curves in Fig.~\ref{lpi_inv_time} to earlier times. In the case of $(r_*/M,n_{\infty}^{i^+})=(907, -5.058)$ 
the curve has enough to make the limit be reasonably close to $-5$, and we conclude that that would also be the case for more distant observer, if we integrate longer in time.

\begin{figure}[ht]
\includegraphics[width=0.5\columnwidth]{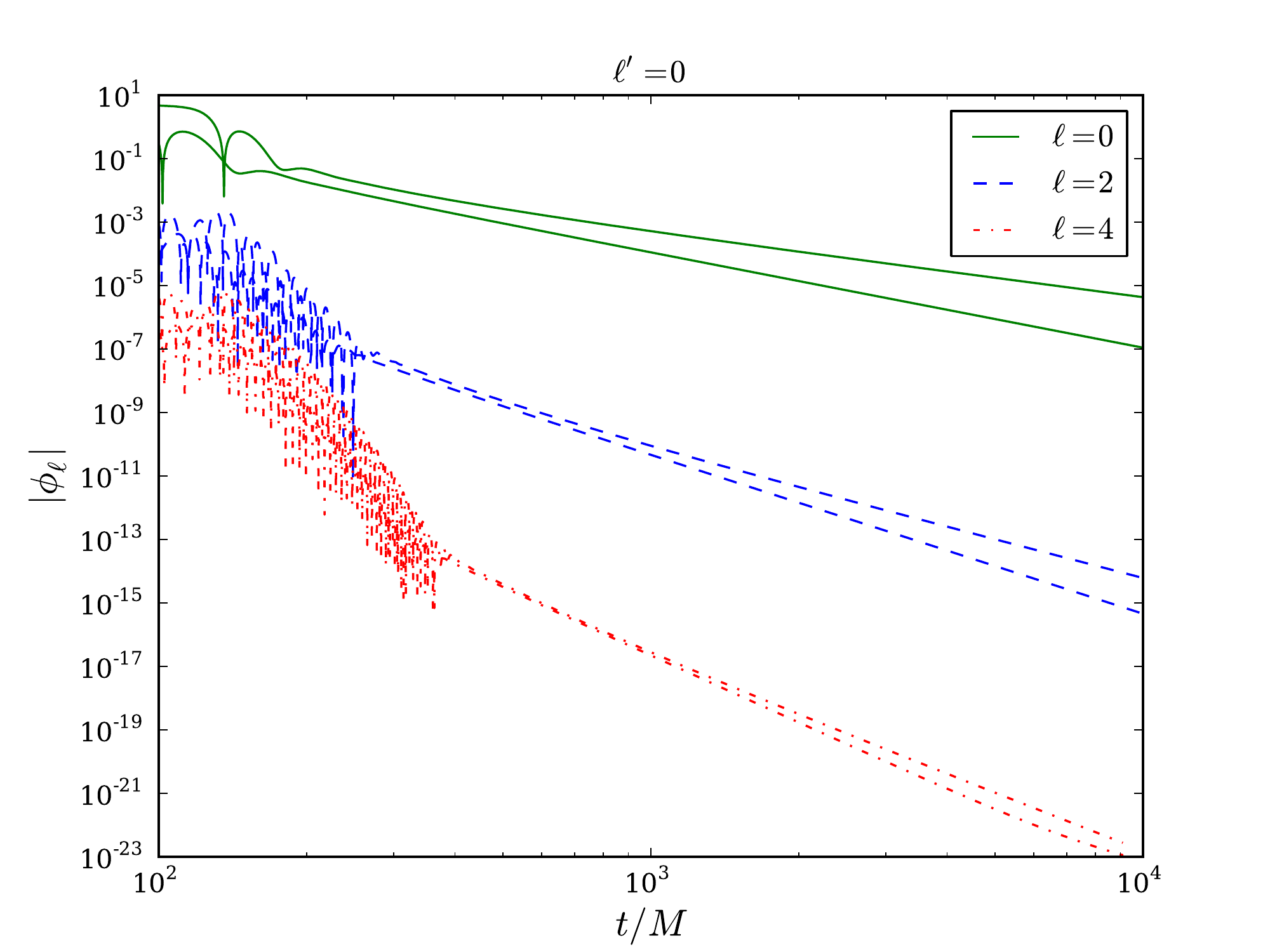}
\includegraphics[width=0.5\columnwidth]{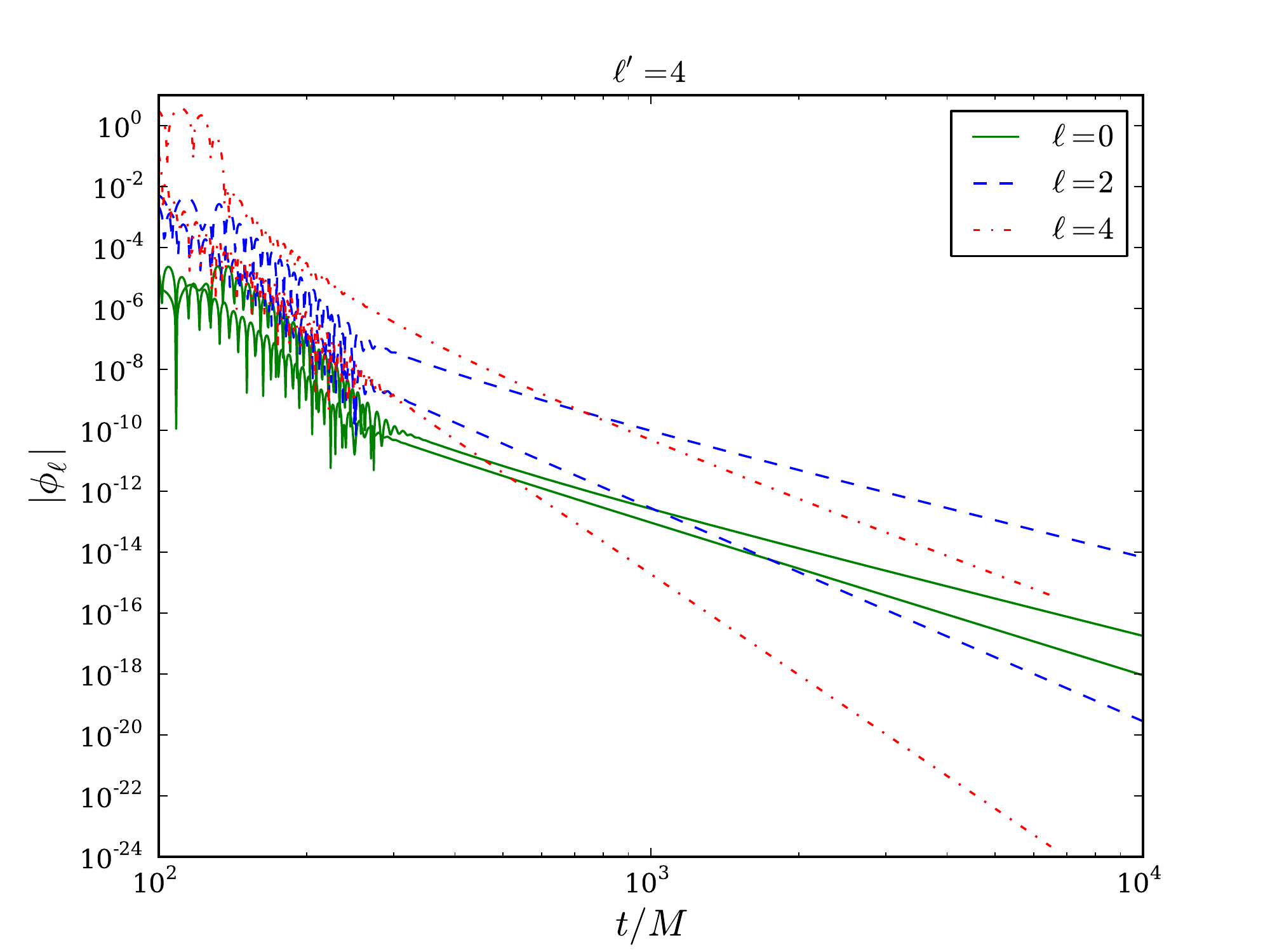}
\caption{Absolute values of the scalar field for the evolution measured by observers  
along null infinity and at $r_*=200M$ for each projected mode as functions of the time, for initial modes $\ell'=0$ (left panel) 
and $\ell'=4$ (right panel). For each case the steeper curve at late times corresponds to the field for the stationary observer at a finite distance. 
The lowest excited mode is the 
dominant mode. It is larger at all times by many orders of magnitude for $\ell'=0$ whereas 
for $\ell'=4$ the lowest excited mode dominates only at late times (see Fig.~\ref{Relative}
for the ratio between the two lowest excited modes). The transition is 
observed as an intermediate `splitting' in the full field.
\label{AbsVals}}
\end{figure}

It may seem puzzling that the local decay rates do not present any `splitting' behavior 
(Fig.~\ref{ProjRates}) whereas the full field does (Fig.~\ref{lpi_inv_time}). This puzzle is 
explained by considering the relative strengths of modes generated due to mode coupling
in Kerr spacetime. We plot in Fig.~\ref{AbsVals} the absolute values of the projected modes 
for initial data with $\ell'=0$ on the left and $\ell'=4$ on the right panel for comparison. 
The values are measured by two observers for each mode: one at null infinity, the other at $200M$. 
The dominant modes for $\ell'=0$ are by many orders of magnitude larger than the up-modes 
generated by mode coupling. This implies that the full field clearly decays with the rate of 
the dominant mode, and no `splitting' occurs. This clean separation of absolute values in 
projected modes is no more valid for initial data with $\ell'=4$ as seen in the right 
panel of Fig.~\ref{AbsVals}. The initial mode starts strongest, but decays fast leaving the 
generated lower modes behind. The null infinity decay rates of the lowest two modes are 
the same, implying that there is no transition of strength between them (compare also 
Fig.~\ref{ProjRates}). The finite distance rates of these modes, however, are different. 
Consequently, the lowest mode with slowest decay dominates at late times, but at early 
times the decay of the faster decaying higher mode may still dominate. This leads to 
the apparent `splitting' behavior at far away distances indicated in Fig.~\ref{lpi_inv_time}. 
We note that the same behavior is also found in the case of odd $\ell, \ell'$ modes.  

We argue that full-field `splitting' appears because the simulation has not run far enough 
for far away observers. Longer evolutions would make far observers near ones. Considering 
Fig.~\ref{lpi_inv_time}, we predict that the asymptotic decay rates measured by far observers 
would be numerically close to $-5$ for longer evolutions. Notably, already with the given 
evolution time the extrapolations for the distant observers curve ``up'' in Fig.~\ref{lpi_inv_time}, 
in support of our prediction. 

\begin{figure}[t]
\center
\includegraphics[width=0.5\columnwidth]{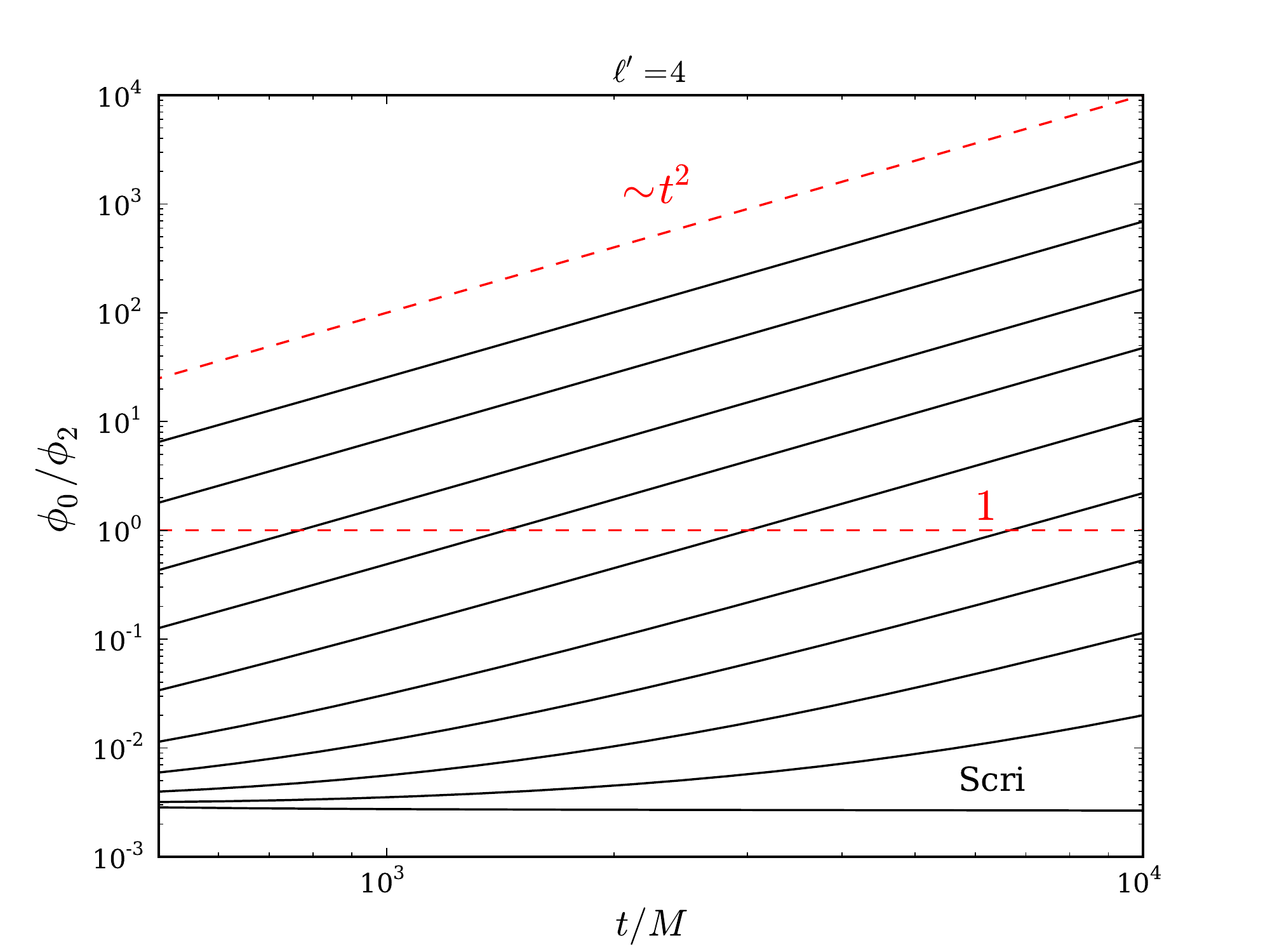}\hfill
\caption{Ratio of the lowest generated modes as a function of time. We plot the ratios of $\ell=0$ 
to $\ell=2$ modes for $\ell'=4$ initial data measured by observers ranging from the vicinity 
of the horizon to null infinity (from top to bottom in the figure). Two dashed red reference lines indicate a ratio
of unity, and the time development of the ratio as $t^2$ between the modes valid for observers close to the
black hole. The plot shows that the lowest mode will dominate for any finite
distance observer at sufficiently late times, but the notion of near and far observers 
depends on the evolution time. There is no transition of decay rates along null infinity, referred to as Scri in the 
above figure.
\label{Relative}}
\end{figure}

Further support for this argument comes from the ratio of the two lowest generated modes, $\phi_{\ell'''=0}/\phi_{\ell''=2}$ for $\ell'=4$, plotted in Fig.~\ref{Relative}. 
This plot gives additional information to the evolution of modes whose decay rates are depicted in 
Fig.~\ref{ProjRates}. As can be inferred from Fig.~\ref{ProjRates}, the rates along null infinity between the
two lowest generated modes are the same. Therefore, their ratio stays below unity and no transition is observed
for the decay along null infinity. For observers closer to the black
hole (upper curves in the figure), however, the ratio evolves in time and goes above unity eventually. The dashed
red line depicting a ratio of 1 can be seen as describing the
transition of the decay rate from the rate of the higher mode to the lowest one. 
The fact that the ratio goes above unity for finite distance observers indicates that the lowest mode becomes
the dominant mode whose decay is observed. The farther away the observer the longer it takes for the lowest
generated mode to dominate. In that sense the `splitting' of the decay rates for the full field is only an 
intermediate behavior due to the relative strength of generated modes. For observers near the black hole, the
difference in the decay rates is $2$, so the ratio follows the $t^2$ curve depicted by the second dashed red line
in Fig.~\ref{Relative}.

\subsection{Near-field}
\label{sec:nh}
\begin{figure}[ht]
\center
\includegraphics[width=0.5\columnwidth]{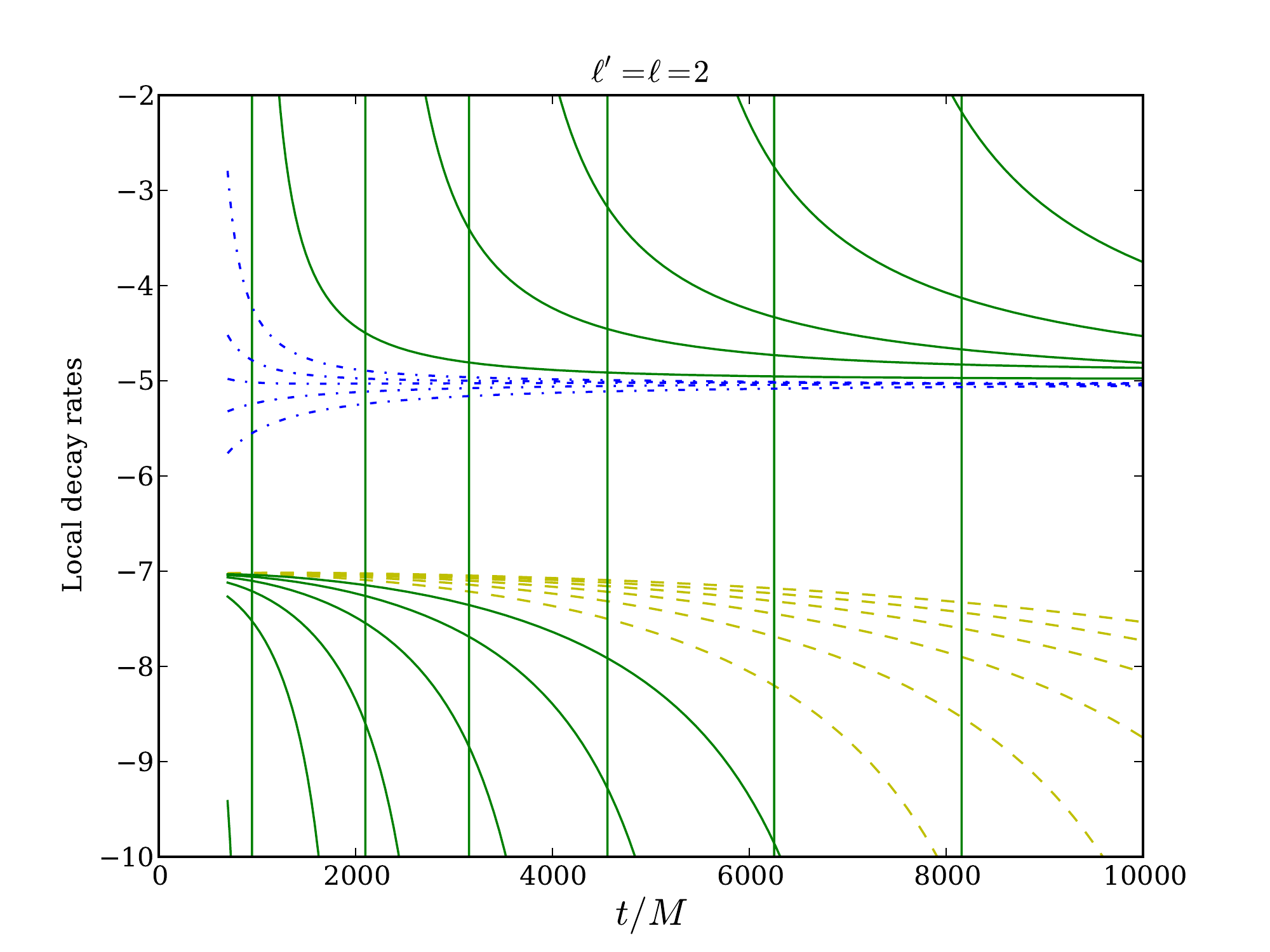}\hfill
\includegraphics[width=0.5\columnwidth]{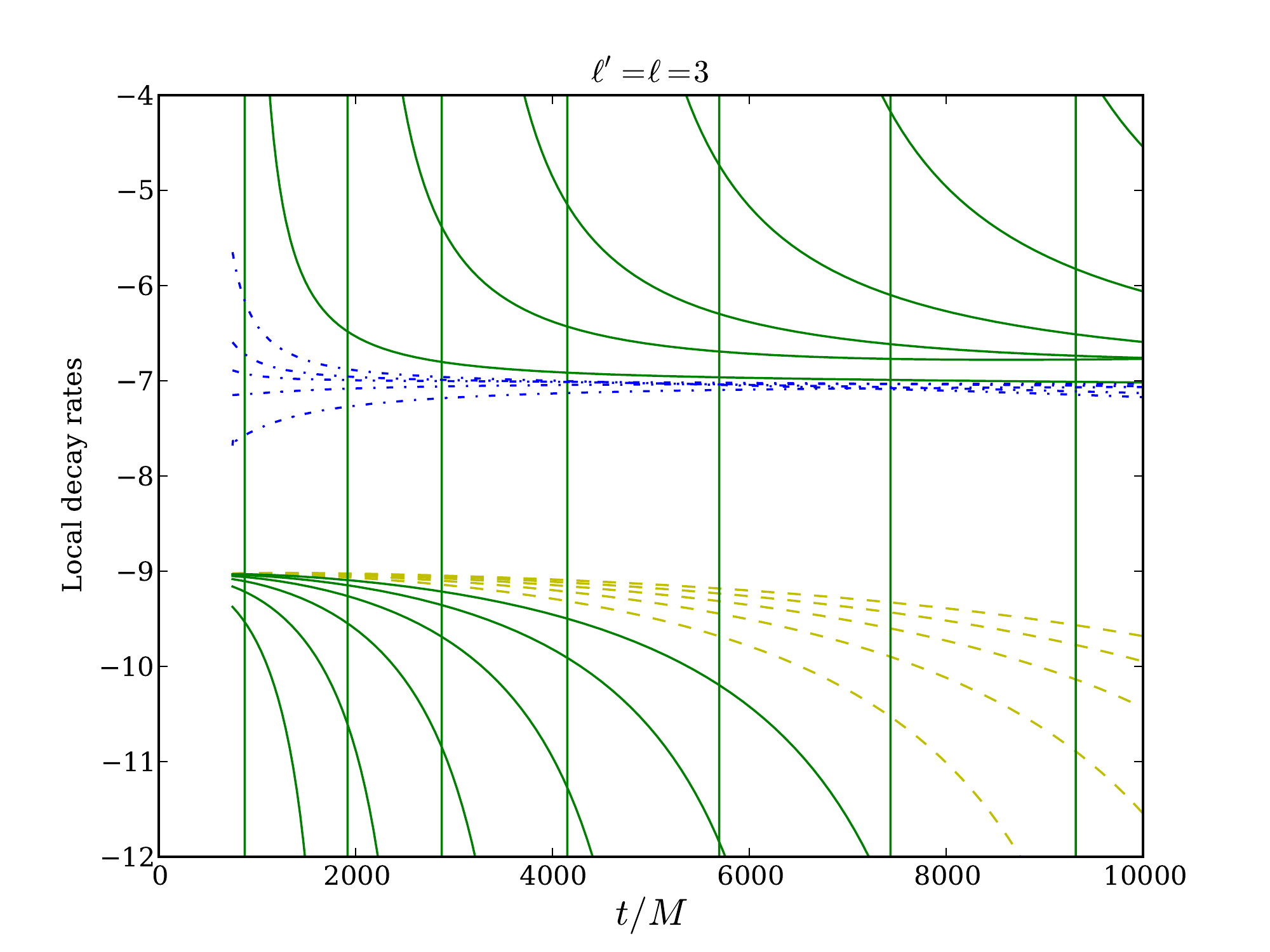}
\caption{Local decay rates for near-horizon observers with $\ell'=\ell=2$ 
(left panel) and $\ell'=\ell=3$ (right panel). The observers 
are located (from large distances to lower distances) at $r_*/M=20,18,16,14,12,10$ 
(yellow dashed), $r_*/M=8,6,4,2,0,-4$ (green solid) and $r_*/M=-8,-12,-20,-70, -\infty$ 
(blue dotted).  
The yellow dashed lines are consistent with Fig.~\ref{ProjRates} and Tab.~\ref{tab:rates}. 
The green solid vertical lines indicate the transition during which the field changes sign. 
The blue dotted lines are the near-horizon rates. The plot indicates that at asymptotically 
late times the near-horizon rates (blue dotted) will dominate for far away observers (yellow dashed) 
after a sign change (green solid).
\label{NHSplit}}
\end{figure}

The picture of decay rates in certain projected modes presented in Fig.~\ref{ProjRates} 
and Tab.~\ref{tab:rates} is modified when near horizon rates are taken into account. 
R\'{a}cz and T\'oth report `splitting' in \cite{Racz:2011qu}\footnote{We thank G\'abor Zs 
T\'oth for pointing out that `splitting' occurs near the horizon for the projected modes.} 
in the projected modes for certain cases. We plot the near horizon 
decay rates for two cases, $\ell'=\ell=2$ and $\ell'=\ell=3$ in Fig.~\ref{NHSplit} on the left and 
the right panels respectively. The observers are located (from large distances to lower distances) at 
$r_*/M=20,18,16,14,12,10$ (yellow dashed), $r_*/M=8,6,4,2,0,-4$ (green solid) and 
$r_*/M=-8,-12,-20,-70, -\infty$ (blue dotted). 

In such cases with $\ell'=\ell$, the determination of the asymptotic decay
rate is particularly difficult due to the near-field intermediate behavior.
Decay rates along $r_*\geq 10M$ denoted by the yellow dashed lines are consistent with 
Fig.~\ref{ProjRates} and Tab.~\ref{tab:rates}. The local rate curves bend down,
however, in preparation for a sign change, which happens along the green solid
lines in the range $r_*\in [-4M,8M]$. The rate after the sign change approaches the same
rate as the blue dotted lines, plotted for observers with $r_*\leq -8M$. The green vertical
lines in Fig.~\ref{NHSplit} indicating sign change can be regarded as the boundary of
`splitting' between the two finite distance decay rates. 

The location dependence observed in Fig.~\ref{NHSplit} provides evidence for the difference between 
intermediate and asymptotic behavior. We suggest that, asymptotically in time, the decay 
of the blue dotted lines in Fig.~\ref{NHSplit} will dominate for all finite observers 
but at any given late time there will be some far away observers that see 
the decay of the yellow dashed lines consistent with Tab.~\ref{tab:rates}.

\begin{figure}[ht]
\center
\includegraphics[width=0.5\columnwidth]{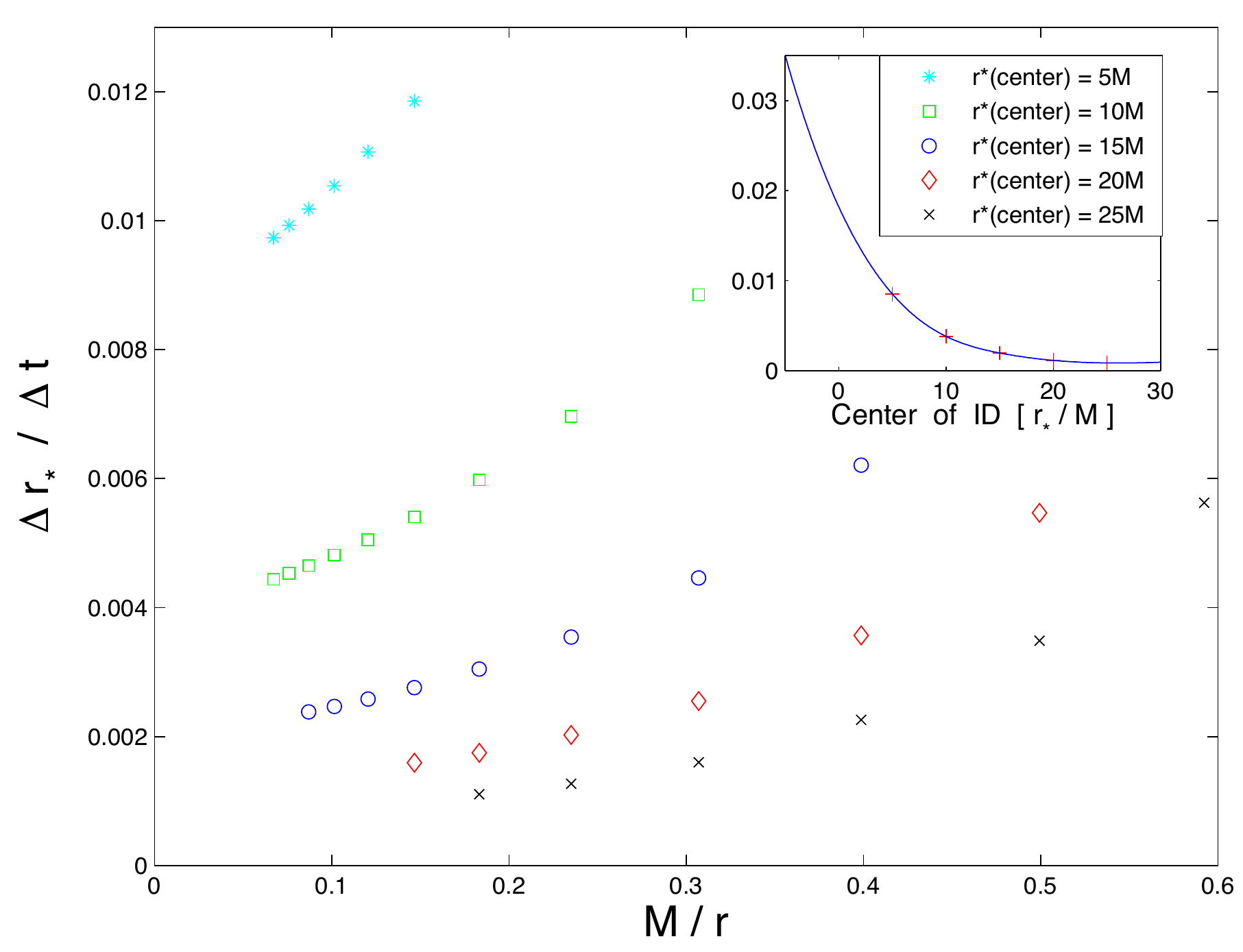}
\caption{Rate at which the boundary of the `splitting' behavior moves as a 
function of the inverse distance of the observer. We show the rate 
for five sets of initial data characterized by the $r_*$ value of the center 
of the initial wave packet. Specifically, we choose $\ell'=2=\ell$ and the 
centers of the initial wave packets are  at $r_{*0}/M=5,10,15,20$ and $25$. 
The insert shows the extrapolated value of the boundary speed at infinity 
($M/r\to 0$) as a function of the location of the center of the initial 
wave packet and a spline extrapolation thereof. 
\label{Speed}}
\end{figure}

Given that the asymptotic behavior dominates for different observers at different times,
a natural question is how the boundary between the two regimes (intermediate and asymptotic) 
propagates to infinity. We see from Fig.~\ref{NHSplit} that the boundary can be taken as where 
the sign of the field changes, depicted by the vertical green lines. The time difference between the vertical 
green lines seems to become larger for the farther away observers, indicating that the speed of transition 
slows down. We would like to know quantitatively, for a given observer, the speed of the transition. 
In Fig.~\ref{Speed}, we plot the rate at which this transition event moves as a function of the 
inverse distance to the observer for the case $\ell'=2=\ell$. We show the rate for five sets of initial 
data characterized by the $r_*$ value of the center of the initial wave packet. Specifically, we set 
the centers of the initial wave packets as $r_{*0}/M=5,10,15,20$ and $25$. 

Near observers see a faster propagation speed of the boundary than distant observers. 
In addition, initial wave packets closer to the black hole cause a faster transition. Both properties
indicate that the transition is a strong field behavior. Intuitively it makes sense that the asymptotic 
decay rate is obtained faster for all observers when the initial
perturbation is in the strong field domain close to the black hole. 
 
The insert shows the extrapolated value of the boundary speed at infinity 
($M/r\to 0$) as a function of the location of the center of the initial 
wave packet and a spline extrapolation thereof. 
Even for far observers, the transition speed seems to approach a non-vanishing limit.
This dependence of the `splitting' behavior on the location of the initial perturbation
has not been observed before.

In conclusion, we distinguish late time behavior for large $t$ 
from asymptotic behavior for $t\to\infty$. In that sense, {\em `splitting' 
in the projected modes should be understood as an intermediate late time 
behavior, and not an asymptotic behavior.} 

Our explanation for this behavior is that successive mode excitations affect 
the decay rate of the late-time tail via inter-mode coupling. For example, the decay rate of 
the $\ell=4$ mode given $\ell'=2$ initial data progresses in multiple channels. 
The direct channel is the excitation of the $\ell=4$ mode by the $\ell'=2$ initial data. 
This direct channel leads to a tail with asymptotic decay rate of $t^{n_0}$ 
where $-n_0=\ell'+\ell+3=2+4+3=9$.\footnote{In what follows we omit the subscript from the asymptotic decay rate $n_{\infty}$ as it is implied.} 
In an indirect channel, the $\ell''=0$ is excited. The excited 
monopole field then acts as a source and creates a second $\ell=4$ tail whose asymptotic  
decay rate is riven by $t^{n_1}$ where $-n_1=\ell''+\ell+3=0+4+3=7$. Similarly, the decay 
rate of the $\ell=2$ mode given  $\ell'=2$ initial data progresses in multiple channels. 
In the direct channel, the $\ell=2$ mode evolves without coupling, 
for which the asymptotic decay rate is given by $t^{n_0}$ where $-n_0=\ell'+\ell+3=2+2+3=7$. 
The indirect channel involves the excitation of the  $\ell''=0$ mode, which then acts as a 
source field for a second $\ell=2$ field whose tail decays asymptotically according to $t^{n_1}$ where 
$-n_1=\ell''+\ell+3=0+2+3=5$. We propose that the reason why in \cite{Racz:2011qu} these values 
(and others) were indeterminate is that the multiple channels could not be computed.  
Note that these observations suggest that the tail decay rates found by Hod 
\cite{Hod:1999rx} are not asymptotic, because they ignore the inter-mode interaction. 
This reasoning also explains the $m$-dependence of the 
`splitting' as observed in \cite{Racz:2011qu,Jasiulek:2011ce}. We summarize our findings in 
Tab.~\ref{tab:rates1}. Notice how these values differ from their counterparts in 
Tab.~\ref{tab:rates}. This intricate inter--mode coupling mechanism has not been suggested before, 
and warrants further study. 

\begin{table}[h]
  \begin{center}
      \begin{tabular}{|c||c|c|c|}
	\hline
        $\ell'$   &   $\ell=0$  &    $\ell=2$   &   $\ell=4$ \\
        \hline\hline 
        0 &  -2 \quad -3   &   -4 \quad -5   &  -6 \quad -7   \\
        2 &  -2 \quad -3   &   -4 \quad -5   &  -6 \quad -7   \\
        4 &  -4 \quad -5   &   -4 \quad -7   &  -6 \quad -9  \\
	\hline
   \end{tabular}\qquad
      \begin{tabular}{|c||c|c|c|}
	\hline
        $\ell'$   &   $\ell=1$  &    $\ell=3$   &   $\ell=5$ \\
        \hline\hline 
        1 & -3 \quad -5   &   -5 \quad -7   &  -7 \quad  -9   \\
        3 & -3 \quad -5   &   -5 \quad -7  &  -7  \quad -9   \\
        5 & -5 \quad -7   &   -5 \quad -9   &  -7 \quad -11 \\
	\hline
   \end{tabular}
\end{center}
  \caption{Same as in Table \ref{tab:rates} with the near--field effect of first--order inter-mode coupling.
\label{tab:rates1}}
\end{table}

Based on \cite{Gleiser:2008} we know that starting with a some value of $\ell'$, all 
dynamically allowed lower modes are excited at higher orders in $(a/r)^2$.\footnote{We use 
here the expansion in powers of $(a/r)^2$ not necessarily as an expansion in a small parameter 
as in \cite{Gleiser:2008}, but ronoather as a counting scheme for the orders of the multiple 
channels; however, for small $a/r$ this also explains the smallness of the amplitude of the 
excited source.} Specifically, starting with an $\ell'=4$ mode and considering the 
$\ell=4$ mode, we have the direct channel that leads to decay rate with $t^{n_0}$ where 
$-n_0=\ell'+\ell+3=4+4+3=11$. We also have two indirect channels, one with the excited quadrupole 
mode and the other with the excited monopole mode.  Naive application of our argument above would lead to decay rate with 
$t^{n_1}$ where $-n_1=\ell''+\ell+3=2+4+3=9$, and the latter the decay rate with $t^{n_2}$ where 
$-n_2=\ell'''+\ell+3=0+4+3=7$.  If this reasoning were correct, one would expect the asymptotic decay 
rate to be $t^{-7}$. However, in practice we find the decay rate to be  $t^{-9}$. We propose that this 
different decay rate is found because the produced monopole does not decay at the same rate that 
a monopole present in the initial data does. Indeed, the former decays as  $t^{-3}$ whereas the latter 
decays as  $t^{-5}$. This difference is proposed to be responsible for the faster decay rate we find. 
In the previously discussed case of $\ell'=2$, $\ell=2$ the produced monopole decays at the same rate as  a monopole 
present in the initial data is, and therefore the decay rate obeyed the naive expectations. See the discussion in \cite{Burko:2013bra}.

It is challenging to observe such decay rates numerically due to
the very high order effect that is needed to produce them. For the $\ell'=4$, $\ell=4$ case and counting powers of $(a/r)^2$, the $\ell''=2$ 
mode is excited at $1^{\rm st}$ order, i.e., at $O((a/r)^2)$, and the $\ell'''=0$ mode is excited 
at $2^{\rm nd}$ order, i.e., at $O((a/r)^4)$. Continuing the counting of orders,  $n_1$ is a $2^{\rm nd}$ 
order effect at $O((a/r)^4)$, while $n_2$ is a $4^{\rm th}$ order effect at $O((a/r)^8)$. 
Since each mode is excited with a small amplitude, higher order excitations are expected to be 
produced with very low amplitudes which set a considerable challenge for numerical computations 
because of the challenges involved with amplitudes comparable or smaller than the floating point 
arithmetic level, and the very long evolution times which are needed for very low initial amplitude 
to dominate despite their slower decay rate. 

Numerical evidence for such higher order tails await further studies \cite{Burko:2013bra}. 
In the meantime, one may tentatively speculate on the basis of the above argument and our numerical results 
of Table \ref{tab:rates1} that the successive mode excitations ultimately impact the asymptotic decay rate 
of each and every mode. With the exception of the case in which $\ell'$ is the slowest decaying mode (for 
which case the decay rates are given by $-n=\ell'+\ell+3$), all other modes ---even or odd--- appear to decay 
according to $-n=\ell'+\ell+1$. 

\section{Summary and Conclusions}\label{sec:disc}
We presented a detailed study of `splitting' of local decay rates in Kerr spacetime observed in 
previous work (\cite{Zenginoglu:2009hd,Racz:2011qu,Jasiulek:2011ce,Harms:2013ib}) both in certain projected modes 
and in the full field. Our simulations are about an order of magnitude longer ($\sim 10^4M$) 
than currently published simulations in the literature, 
which allows us to determine power indices that were considered, based on shorter evolutions, as 
ambiguous. The determination of these indices suggests a spin-dependent mechanism by which a 
`splitting'--like behavior in the decay rates is produced.
To achieve long simulation times without
destroying relevant features of the solution due to boundary effects, we attached
hyperboloidal layers both in the positive and the negative 
directions in the tortoise coordinate in Kerr spacetime. This procedure leads to a horizon-penetrating,
hyperboloidal coordinate system, completely removing artificial boundaries from the numerical simulation (see also \cite{Bernuzzi:2012ku} for the construction in Schwarzschild spacetime). 
It is a new result that a hyperboloidal layer can be attached in the strong field 
domain near the future event horizon of Kerr spacetime, which may be of potential usefulness also for other applications. 
Some interesting problems where this method might be used are the study of superradiance (see, for example, recent work \cite{Csizmadia:2012kq}), 
and the instability of perturbations of extremal Kerr spacetimes (\cite{Aretakis:2012ei}).

Having access to long simulations and studying various aspects of the solution from local decay
rates to relative amplitudes of projected modes, we obtained a detailed understanding of the 
mechanisms underlying near-horizon and far-field `splitting.'  
These two aspects of `splitting' seem unrelated to each other. 
`Splitting' near the black hole appears in the projected modes, and can be observed with 
standard codes. It is surprising that it has been discovered only recently. `Splitting' far away appears in 
the full field and is due to the competition between the amplitudes of projected modes. 

In cases where full-field `splitting' appears (for example, Fig.~\ref{lpi_inv_time}) we observe that 
different modes have different decay rates. Eventually the lowest mode dominates because of its 
slowest decay (Fig.~\ref{AbsVals}). This transition happens at different times for different 
observers, which appears as `splitting' in the decay rates of the full field. Asymptotically in time, 
the theoretical decay rates are valid. In that sense, the `splitting' of the full field is only 
an intermediate behavior.

`Splitting' in certain projected modes is of a different nature arising from 
the excitation of lower modes. Here we observe that for any given time there are three 
domains with three different local decay rates (see Figs.~\ref{ProjRates} and \ref{NHSplit}): 
very close to the horizon, far away from the horizon, and near infinity. There is a transition in which the field changes sign and the decay rates approach 
the near-horizon rates. The boundary of this transition moves slowly towards infinity (Fig.~\ref{Speed}), which suggests that asymptotically in time 
near-horizon rates will dominate at all finite distances. For all practical purposes, however, 
`splitting' is a real effect for these modes at any given (late) time. We also argued that successive 
mode excitations affect the decay rate of the late-time tail of all modes. However, the effects of 
lower mode excitations on any given mode's decay rate are difficult to observe due to numerical 
challenges. This is an area certainly requiring further work. 


\section*{Acknowledgments}
We thank G\'abor Zs T\'oth and Istv\'an R\'acz for discussions. AZ is supported by the NSF Grant 
PHY-1068881, and by a Sherman Fairchild Foundation grant to Caltech. GK acknowledges research 
support from NSF Grant Nos. PHY-1016906, PHY-113566 and PHY-1303724. LMB is supported by a NASA EPSCoR RID 
grant and by NSF grants PHY-0757344, PHY-1249302, DUE-0941327 and DUE-1300717. Initial work on this project was 
done while LMB was at the University of Alabama in Huntsville. Most of the 
data presented in this work were generated on the Air Force Research Laboratory CONDOR supercomputer. 
GK also acknowledges support from AFRL under CRADA No. 10-RI-CRADA-09.

\appendix

\bibliographystyle{spphys.bst} 
\bibliography{refs}

\end{document}